\documentclass[graybox]{svmult}

\usepackage{mathptmx}       % selects Times Roman as basic font
\usepackage{helvet}         % selects Helvetica as sans-serif font
\usepackage{courier}        % selects Courier as typewriter font
\usepackage{type1cm}        % activate if the above 3 fonts are
\usepackage{makeidx}         % allows index generation
\usepackage{graphicx}        % standard LaTeX graphics tool
\usepackage{multicol}        % used for the two-column index
\usepackage[bottom]{footmisc}% places footnotes at page bottom

\usepackage{color}

\newfont{\fcm}{cmssdc10 scaled 900}
\let\oldmarginpar\marginpar
\renewcommand\marginpar[1]{\-\oldmarginpar[\raggedleft\footnotesize\color{red} #1\color{black}]%
{\raggedright\footnotesize #1}}

\makeindex             % used for the subject index
\begin{document}

\title*{Ionization Conditions in the Central Giant \textsc{H\,ii} Region of NGC~5253}
\author{A. Monreal-Ibero, J. M. V\'{\i}lchez, J. Walsh and C. Mu\~noz-Tu\~n\'on}
\institute{A. Monreal Ibero,  J. M. V\'{\i}lchez \at IAA-CSIC, Glorieta de la Astronom\'{\i}a, s/n, 18008 Granada, Spain \email{[ami,jvm]@iaa.es}
%\and J. M. V\'{\i}lchez \at IAA-CSIC, Glorieta de la Astronomia, s/n, 18008 Granada, \email{jvm@iaa.es}
\and J. Walsh \at ESO Karl-Schwarzschild-Strasse 2, 85748 Garching bei M\"unchen, Germany \email{jwalsh@eso.org}
\and C. Mu\~noz-Tu\~n\'on \at IAC, c/ V\'{\i}a L\'actea, s/n, 38205, La Laguna, Spain \email{cmt@iac.es}
}
\maketitle

\abstract{We are carrying out a detailed 2D spectroscopic study of the
  central 210~pc$\times$130~pc of the blue compact dwarf (BCD) galaxy NGC~5253. This
  contribution presents the ionization conditions and chemical abundance
    patterns for the different kinematic components of the ionized gas 
detected in our previous work. All the kinematic components
  present an excess in nitrogen abundance. In particular, the broad velocity
  component has larger excess than the narrow one by a factor of
  $\sim1.4\pm0.4$ which is consistent with a scenario for the Giant
  \textsc{H\,ii} Region where the broad component of the emission lines traces
  the material expelled by the two central Super Star Clusters while the
  narrow one is associated to previously existing ionized gas.}

\section{Introduction}
\label{intro}

\begin{figure}[b]
\sidecaption
% For example, with the graphicx style use
\includegraphics[width=0.50\textwidth]{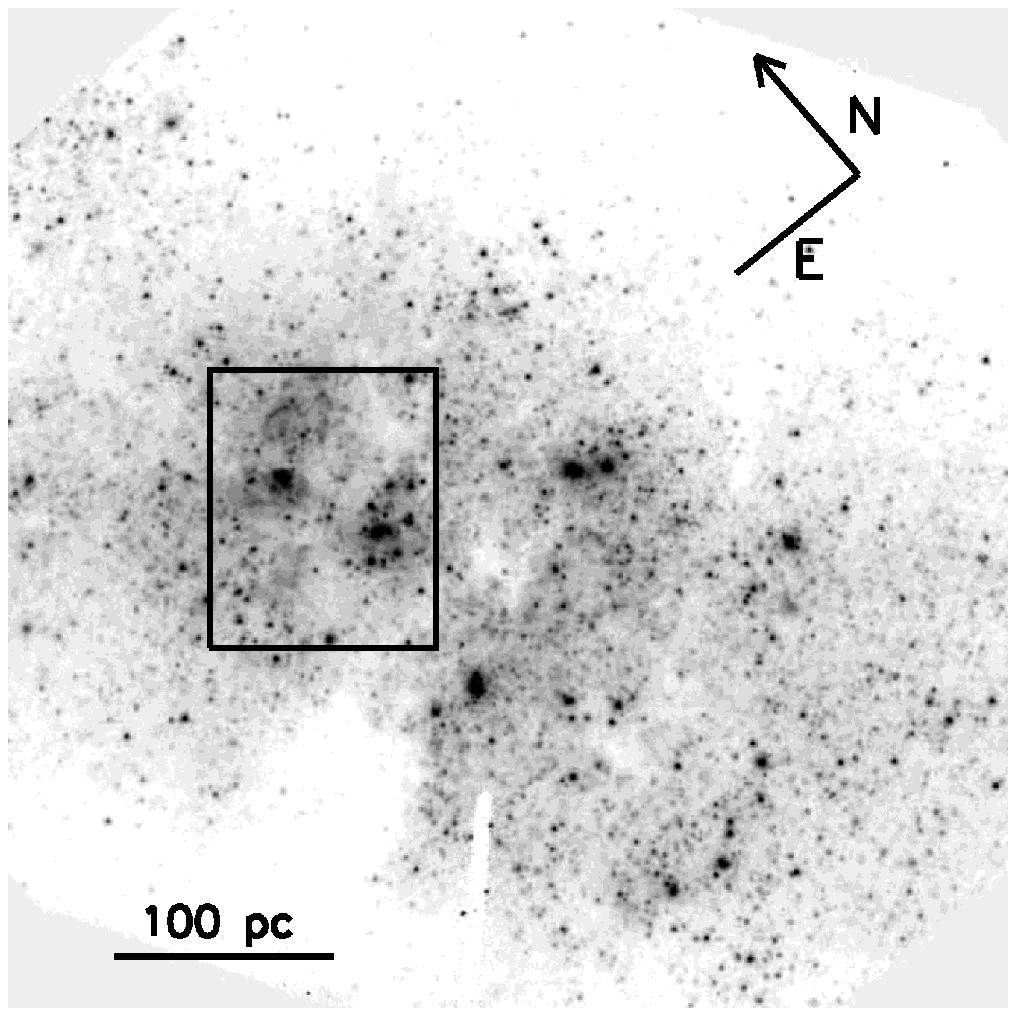}
\hspace{0.4cm}
\includegraphics[width=0.40\textwidth]{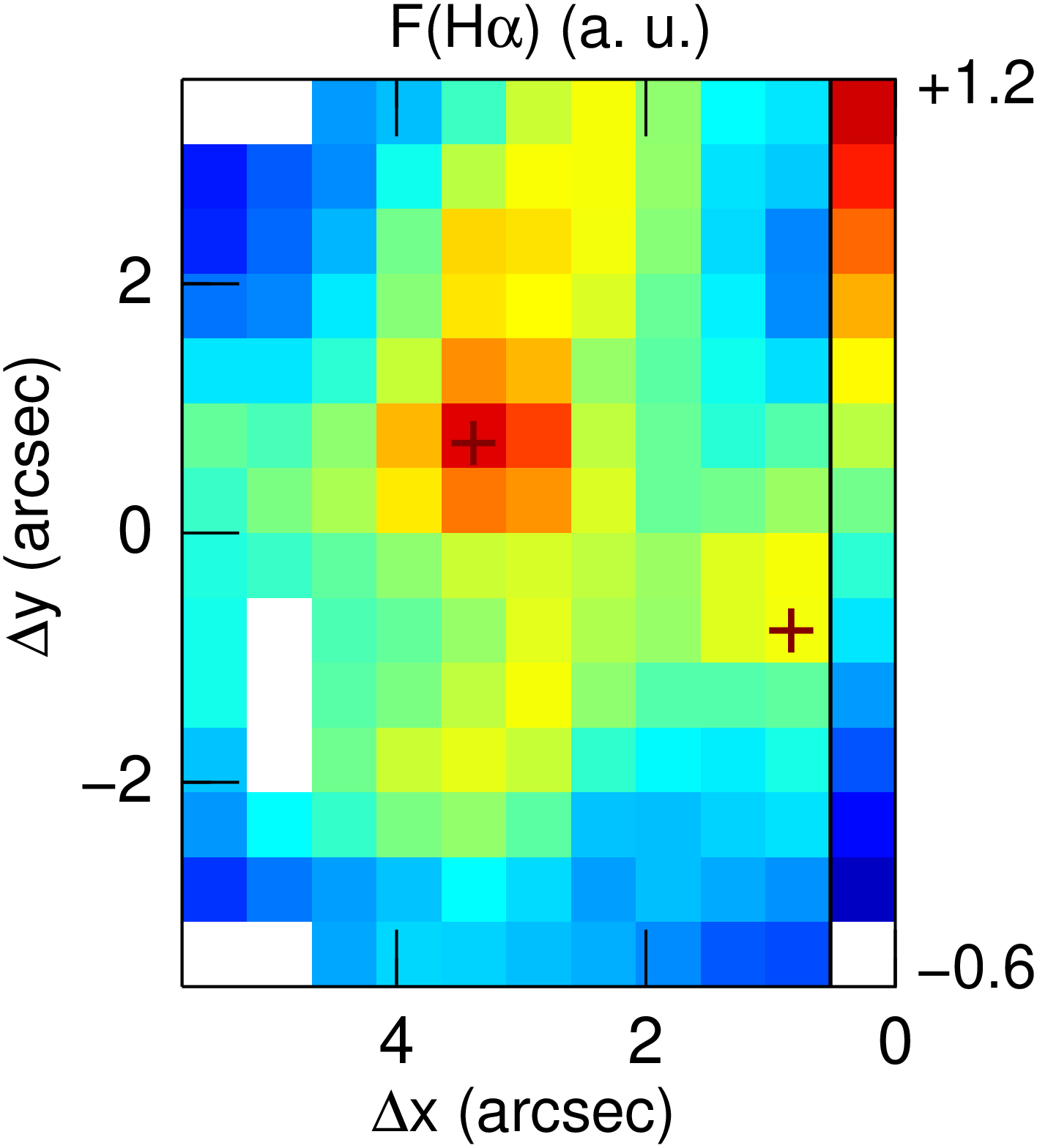}
%
% If no graphics program available, insert a blank space i.e. use
%\picplace{5cm}{2cm} % Give the correct figure height and width in cm
%
\caption{\color{black}\emph{Left:} Gray-scale image made out of the combination of HST-ACS
  images taken with the F435W ($B$), F658N (H$\alpha$), and F814W ($I$)
  filters (programme 10608, P.I. Vacca) with the area analyzed in this
  contribution overplotted as a black rectangle. The orientation and
  spatial scale are indicated. \emph{Right:} Ionized gas distribution
  as traced by the H$\alpha$ emission line. The position of two peaks of
  continuum emission \#1 and \#2 (see \cite{mon10a}) are marked with
  crosses. Map is presented in logarithmic scale in order to emphasize the
  relevant morphological features. Units are arbitrary. We keep the same
  coordinate system as in \cite{mon10a}.}
\label{ngc5253}
\end{figure}

NGC~5253 is a blue compact dwarf (BCD) galaxy located in the Centaurus A/M~83 group
\cite{kar07} which is suffering a burst of star formation probably triggered
by a previous encounter with M~83 \cite{van80}. Because of its proximity
($D=3.8$~Mpc, \cite{sak04}), this galaxy constitutes an optimal target for the
study of the starburst phenomenon. Recently, we performed a detailed study of
NGC~5253, mapping its central part with 2D 
optical spectroscopy \cite{mon10a}. Among other results, we delimited very precisely the area
enriched with extra nitrogen and explored the possibility of Wolf-Rayet stars as the cause of this enrichment.
We also showed that the kinematics of this object is
very complex. In particular, we needed up to three kinematically distinct components to properly reproduce the profile of the brightest emission lines. While two of them, C1 and C3, were relatively narrow, C2 presented supersonic velocities (see \cite{mon10a} for details about the extend of each component and its typical velocity dispersion). These results, together with our findings about the extinction
structure and electron density, were consistent with a scenario in which the two
massive Super Star Clusters (SSCs) at the center of the galaxy are producing
an outflow that encountered previously existing gas.
In this contribution, we
will explore whether the ionization conditions of the individual components
detected in the study of the kinematics also supports this picture. The area
considered in this study is depicted in Fig.~\ref{ngc5253}.

\section{The data}
\label{data}
% Always give a unique label
% and use \ref{<label>} for cross-references
% and \cite{<label>} for bibliographic references
% use \sectionmark{}
% to alter or adjust the section heading in the running head

The data used here were taken at the VLT with the ARGUS mode of FLAMES using
the 0.52$^{\prime\prime}$/lens sampling and the L682.2 and L479.7
gratings. Thus, we could map an area of
$11.5^{\prime\prime}\times7.3^{\prime\prime}$ with a spectral resolution of
$R\sim$12\,500 in an unbiased manner, obtaining information about the main
emission lines in the optical. Seeing was
$\sim0.8^{\prime\prime}-1.0^{\prime\prime}$ allowing for a very good spatial
resolution. Details about data reduction and processing can be found in
\cite{mon10a}. The subject of this analysis (i.e. the Giant \textsc{H\,ii}
Region) occupied roughly the left half of the ARGUS array. Its morphology
is presented in the right part of Fig. \ref{ngc5253}: there is a main peak of
emission and two tongue-shaped extensions towards the north-west and
south-east.
 
\begin{figure}[t]
\sidecaption[t]
\includegraphics[width=0.27\textwidth]{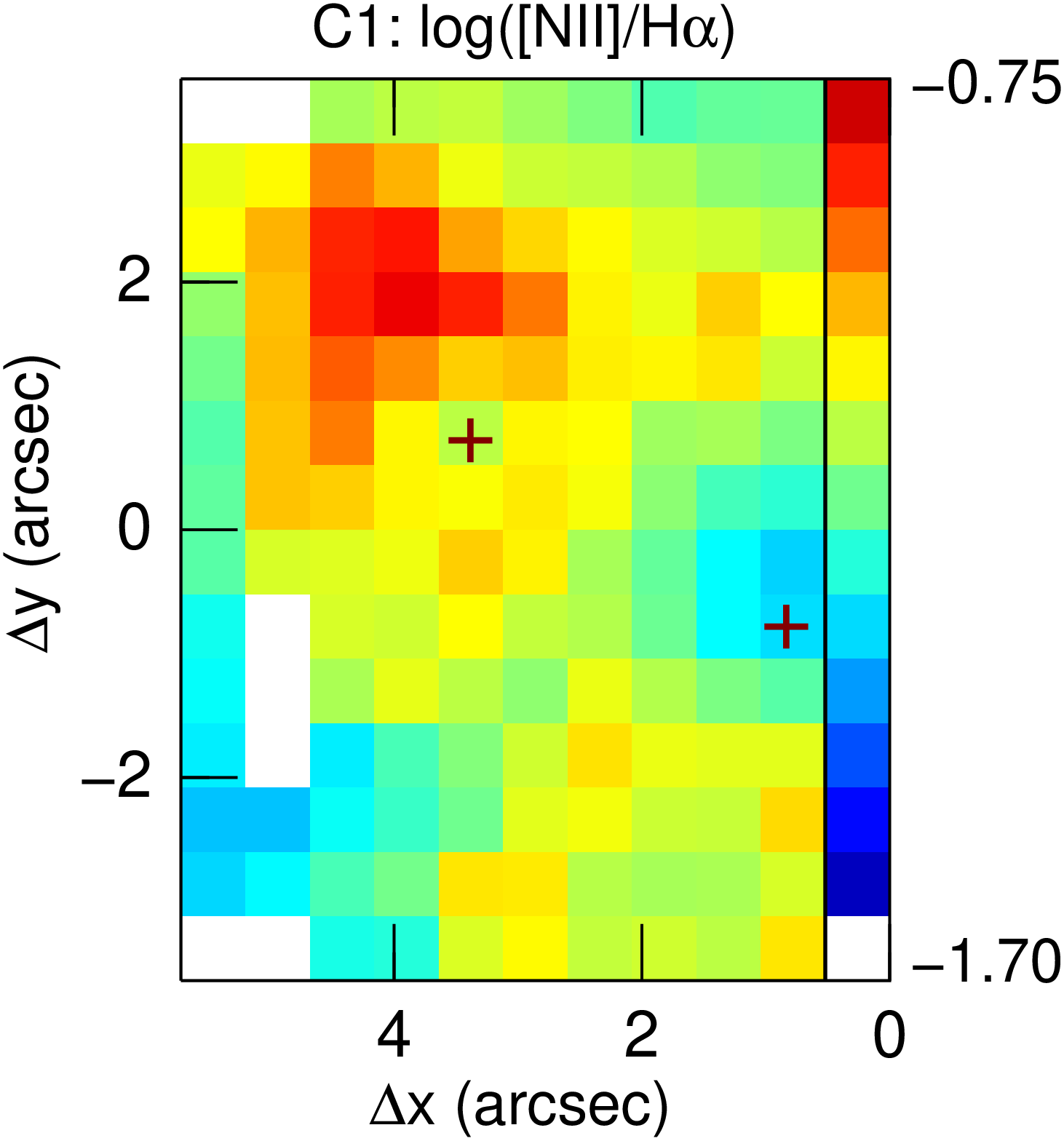}
\includegraphics[width=0.27\textwidth]{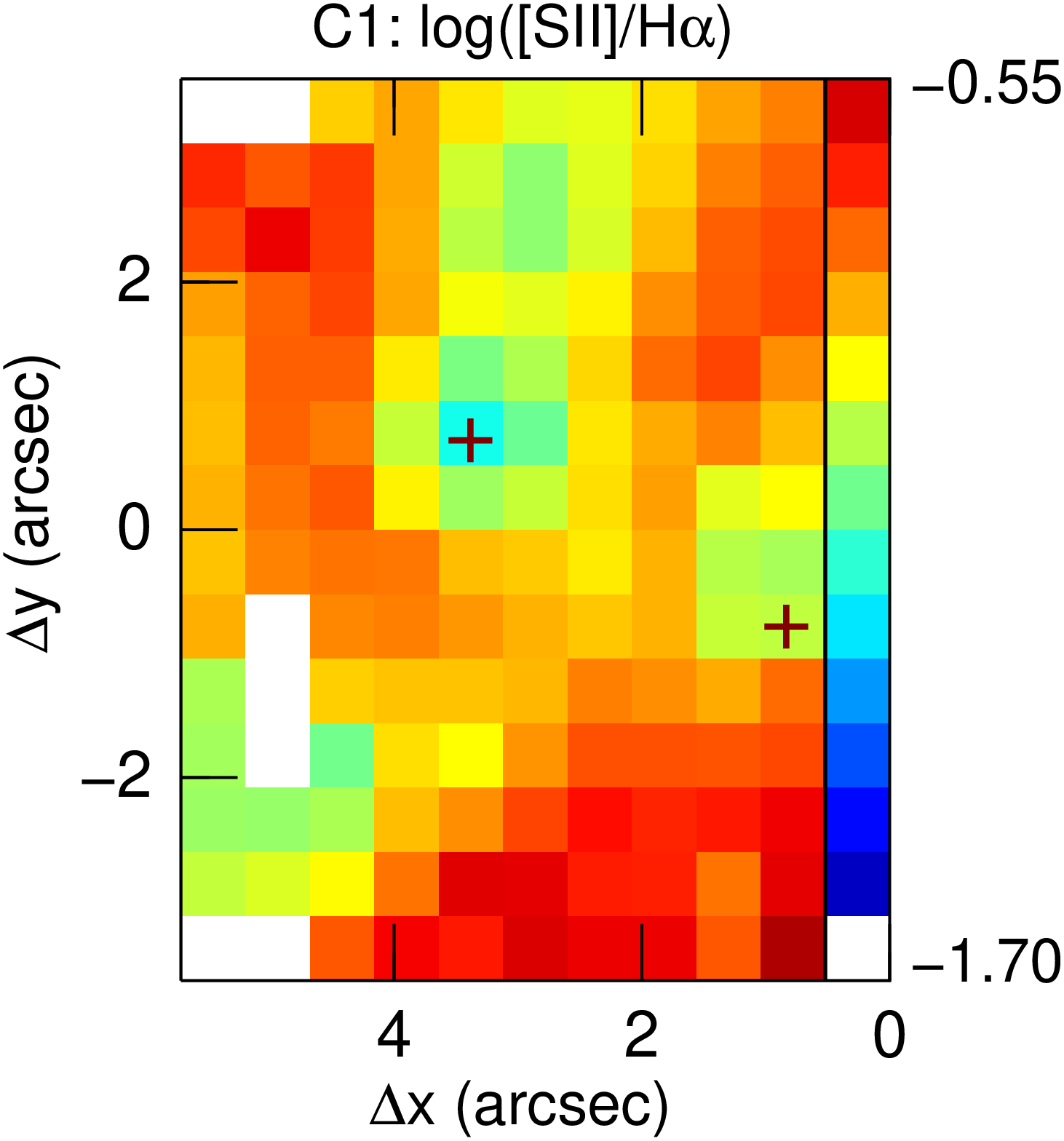}
\includegraphics[width=0.27\textwidth]{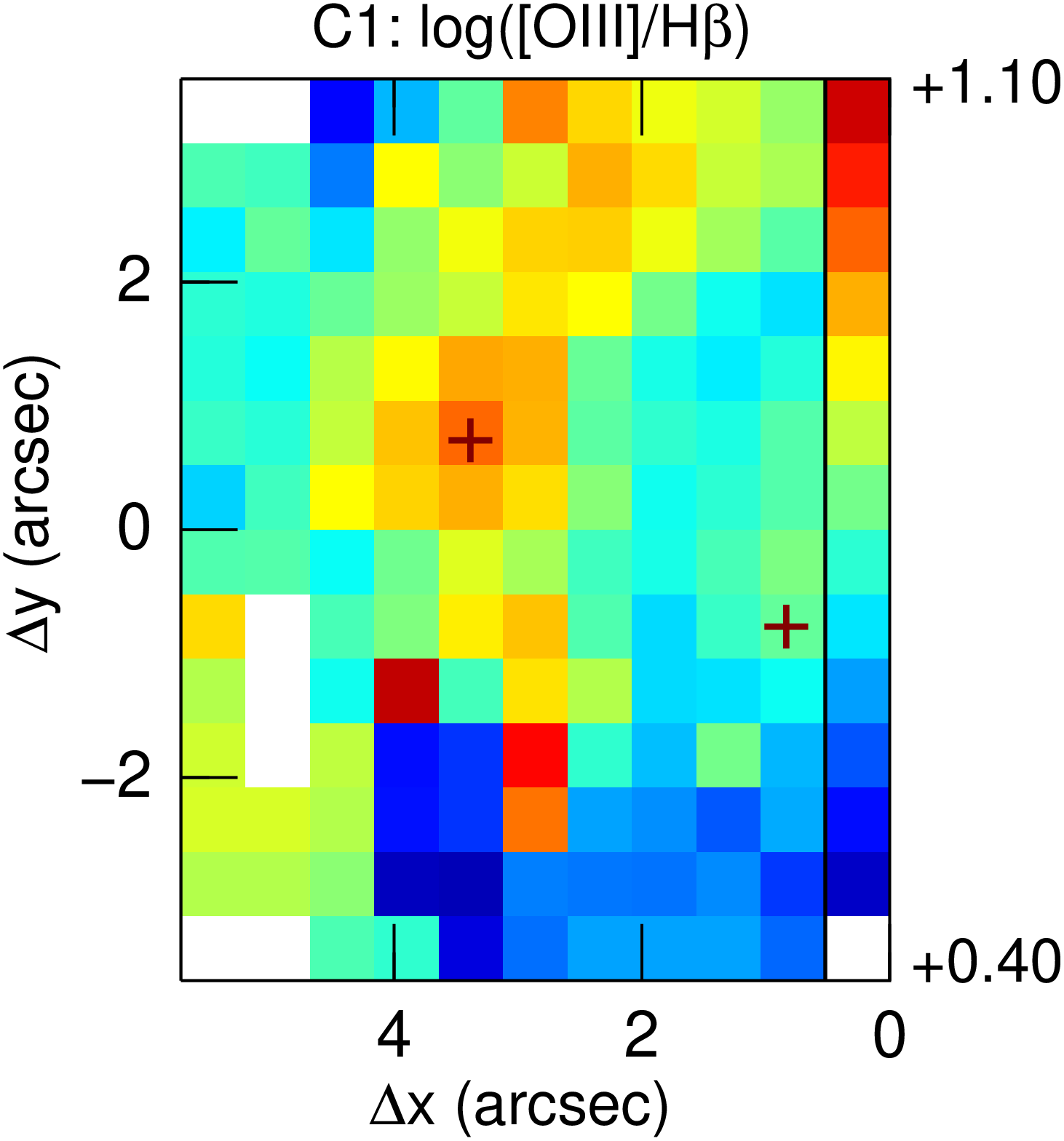}\\
\includegraphics[width=0.27\textwidth]{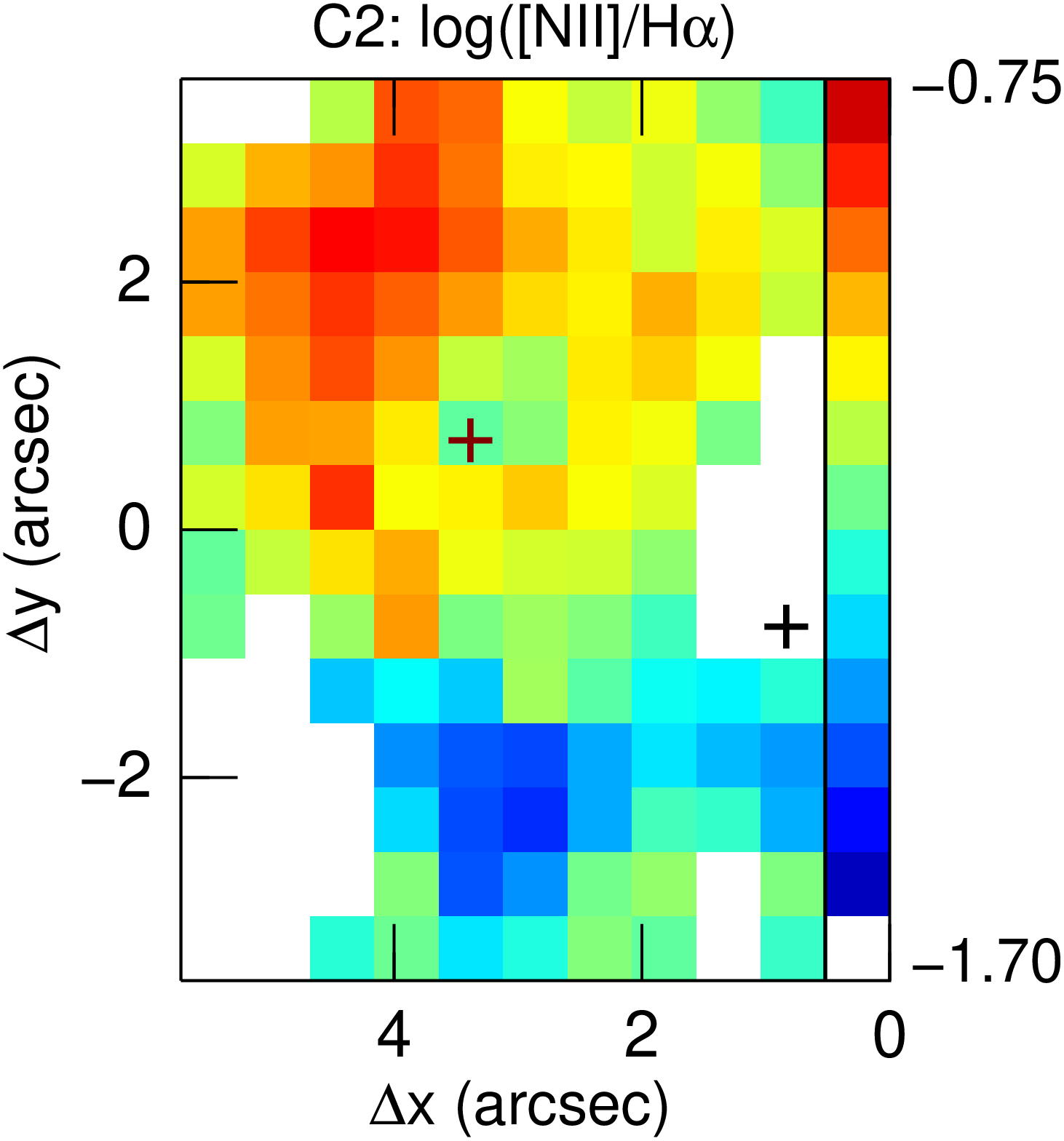}
\includegraphics[width=0.27\textwidth]{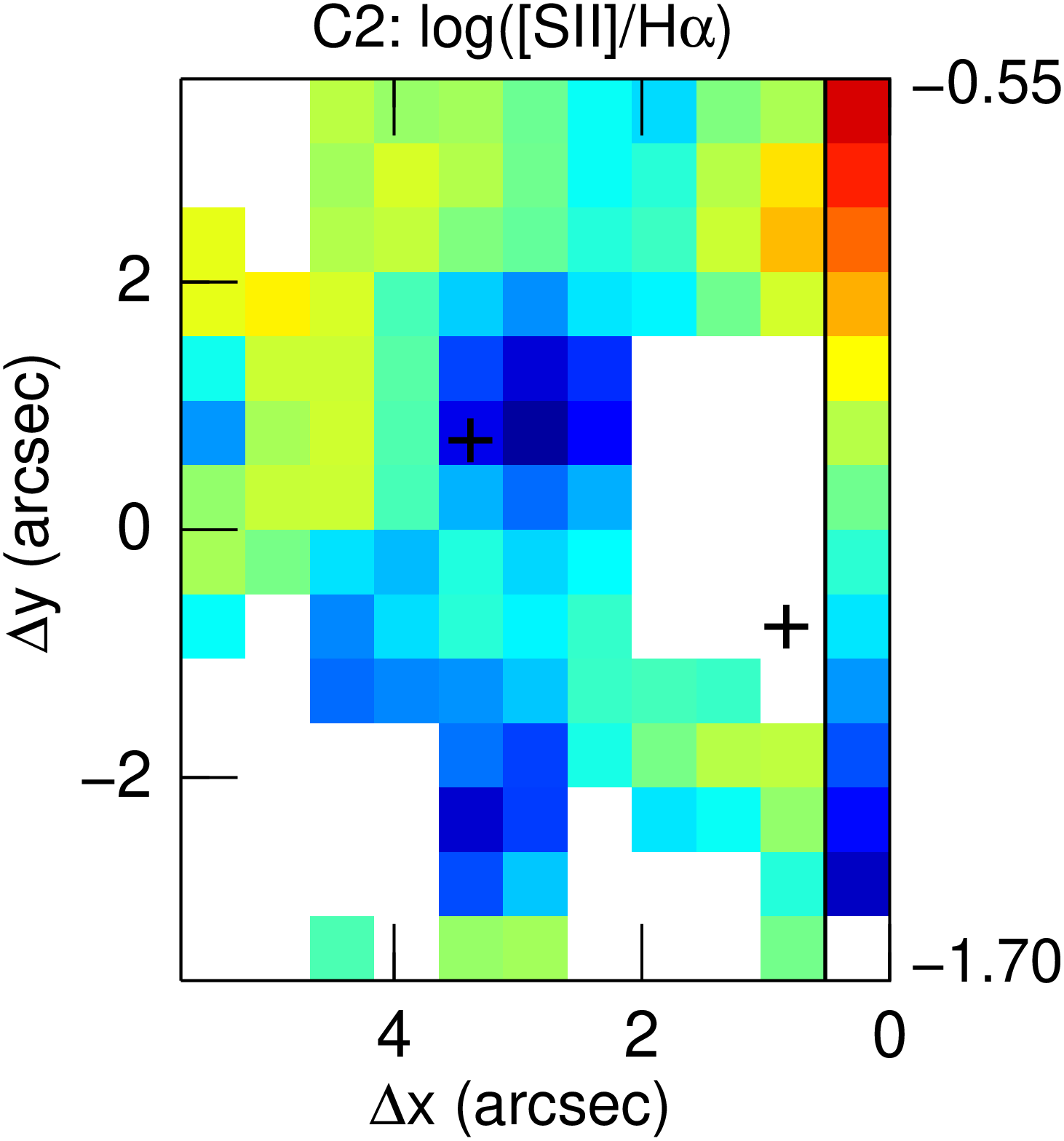}
\includegraphics[width=0.27\textwidth]{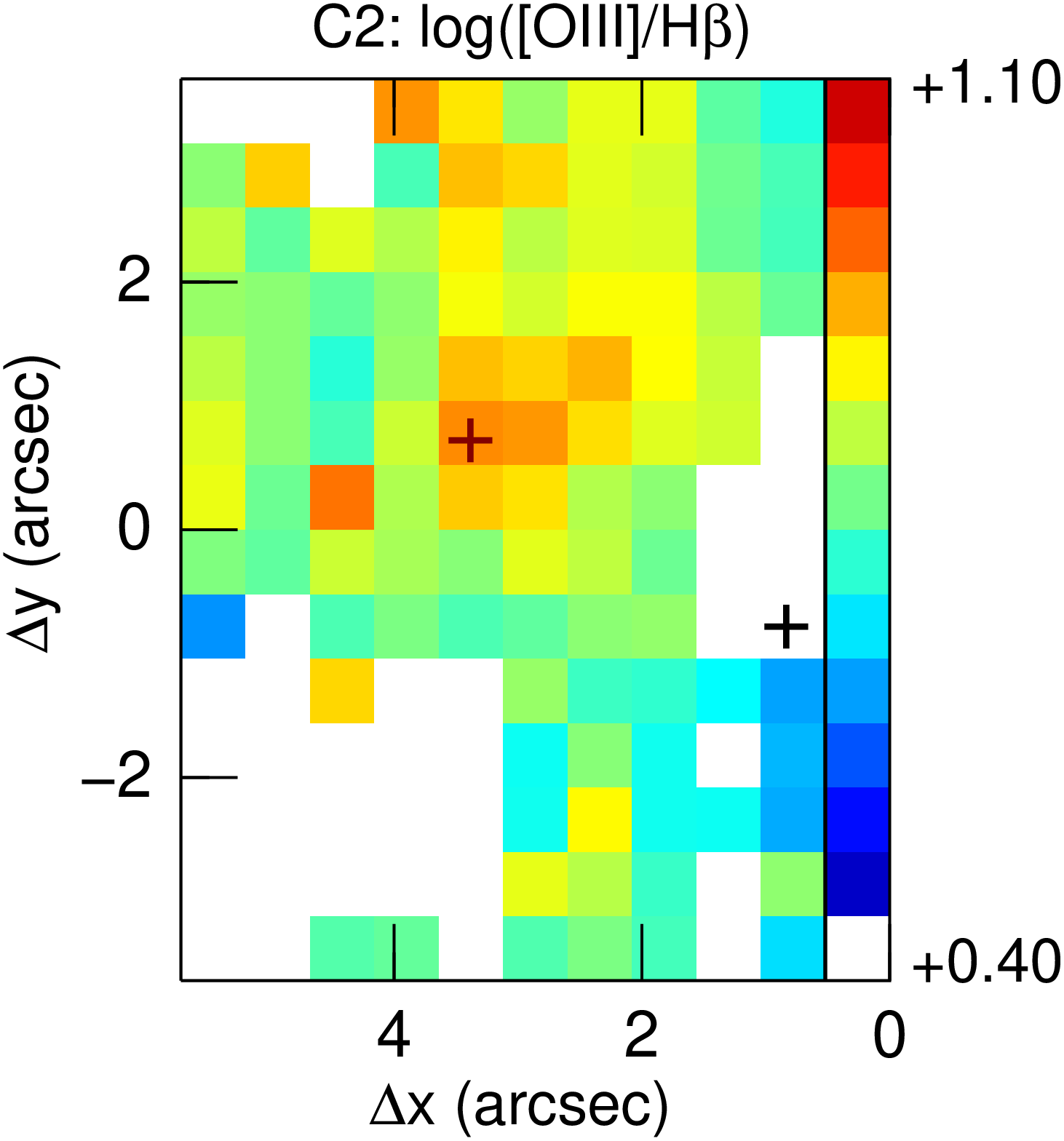}\\
\includegraphics[width=0.27\textwidth]{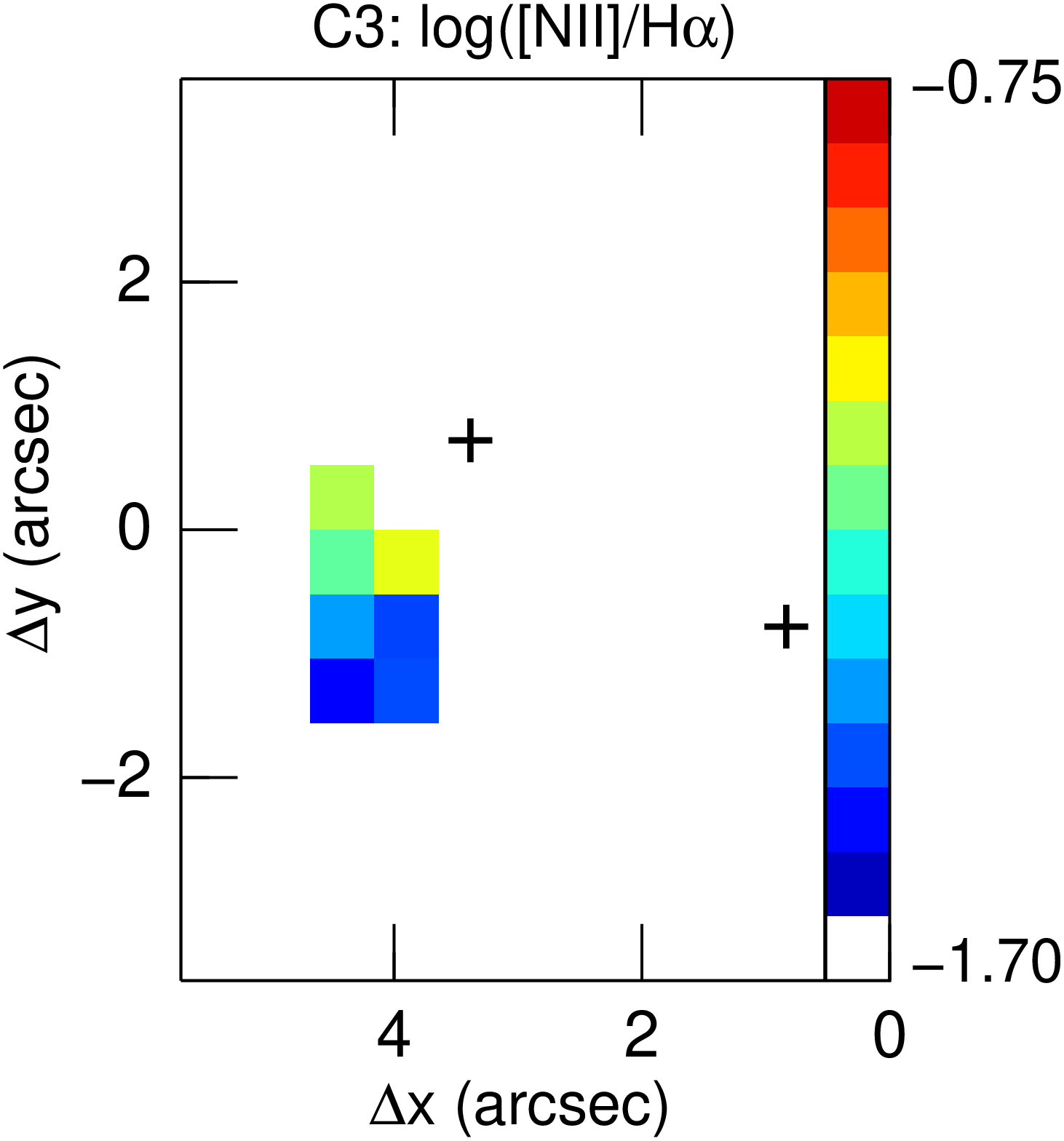}
\includegraphics[width=0.27\textwidth]{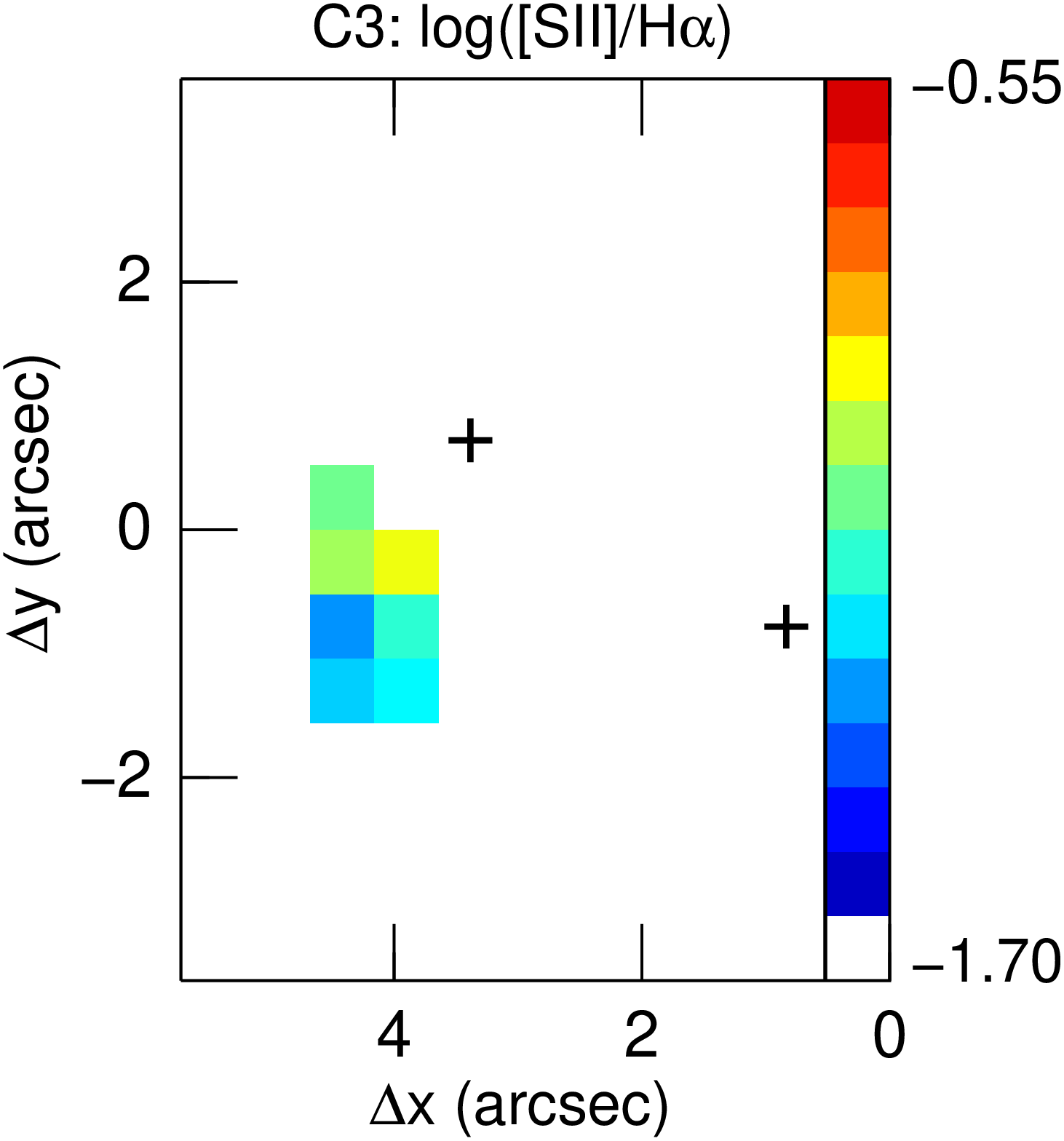}
\includegraphics[width=0.27\textwidth]{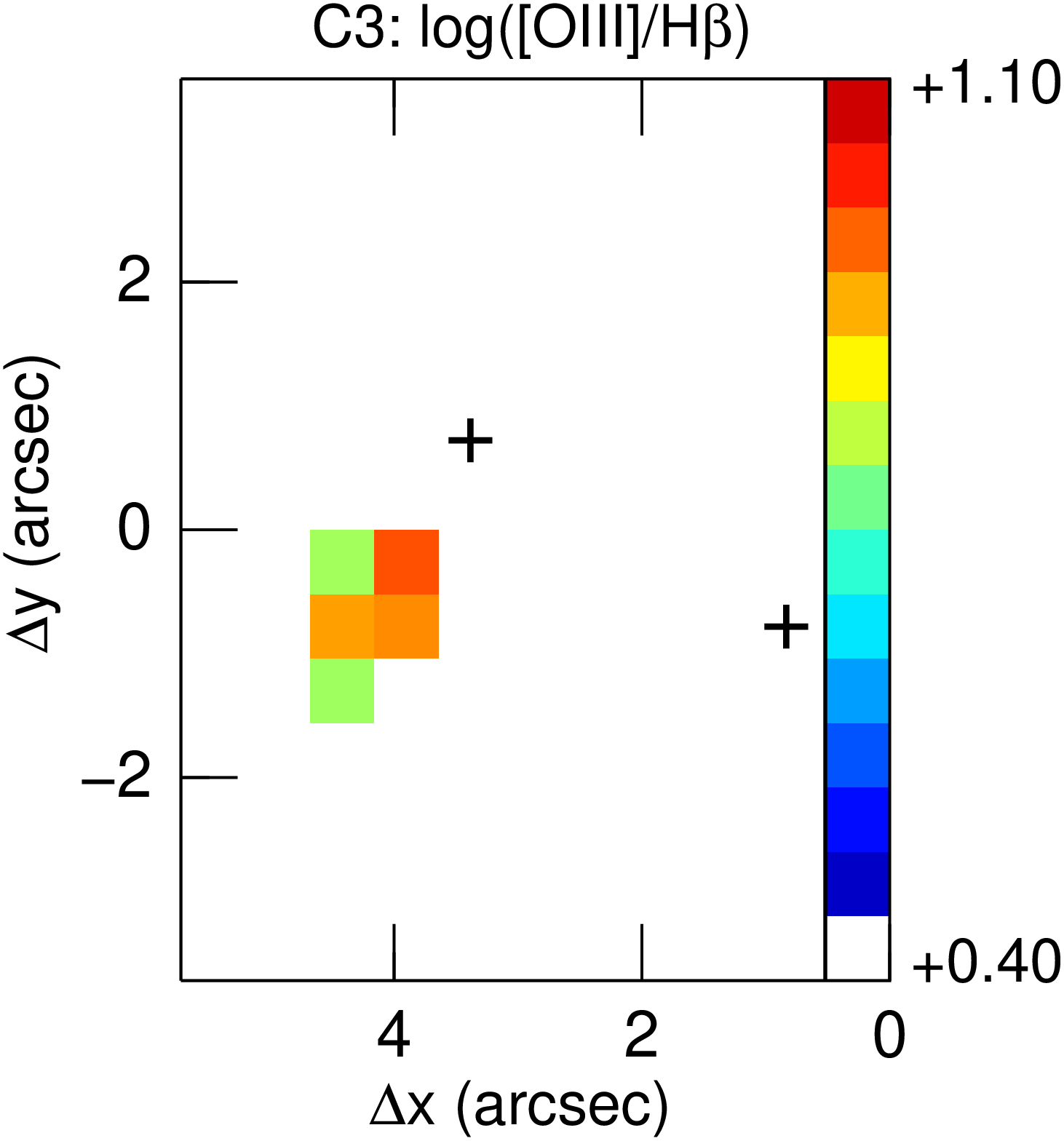}\\
%
% If no graphics program available, insert a blank space i.e. use
%\picplace{5cm}{2cm} % Give the correct figure height and width in cm
\caption{Emission line ratio maps for the three components fitted to the line profile for the kinematic analysis. The position of two peaks of continuum emission \#1 and \#2 (\cite{mon10a}) are marked with crosses. }
\label{mapasdiag}       % Give a unique label
\end{figure}

\begin{figure}[t]
\sidecaption[t]
\includegraphics[width=0.35\textwidth]{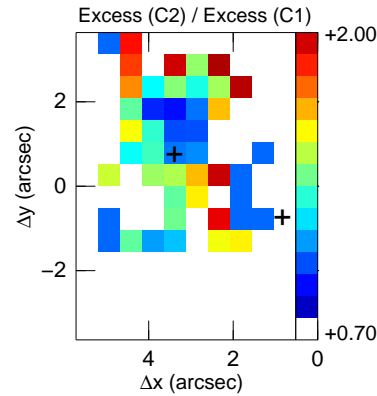}
%
% If no graphics program available, insert a blank space i.e. use
%\picplace{5cm}{2cm} % Give the correct figure height and width in cm
\caption{Ratio between the nitrogen excess determined for the second (i.e. wide) and first (i.e. narrow) component. The position of two peaks of continuum emission \#1 and \#2 (see \cite{mon10a}) are marked with crosses. }
\label{cociexceso}       % Give a unique label
\end{figure}

\section{Results}
\label{resultados}

Maps for the line ratios involved in the so-called BPT diagrams
\cite{bal81,vei87} for the three kinematically distinct components
%\mar{how are these components defined? kinematically? If so, by what $\sigma$? Please briefly explain.} 
are presented in Fig. \ref{mapasdiag}.
For all of them, the \textsc{[O\,iii]}/H$\beta$ and
\textsc{[N\,ii]}/H$\alpha$ maps are relatively similar to those obtained
after fitting each emission line to a single Gaussian. However, the \textsc{[S\,ii]}/H$\alpha$
maps for the narrow component present larger line ratios than for the broad
one. This is exactly what we would expect in a scenario like the one sketched
in Fig. 20 of \cite{mon10a}, where the broad component would be ionized by the
stars in the SSCs while both, shocks and photoionization, could contribute to
the line ratios associated with the shell.

Following the same procedure as in \cite{mon10a}, we estimated the relative
excess in nitrogen.
The three fitted components presented a nitrogen excess. 
Since C3 was detected only in seven spatial elements (\emph{spaxels}) and in an area marginally contaminated, results for this component were uncertain and will not be considered further.
Fig. \ref{cociexceso} presents the ratio between the nitrogen excess measured
for C2 (i.e. broad component) and C1 (i.e. narrow one). Clearly, the broad
component presents more nitrogen excess than the narrow one with a mean
($\pm$r.m.s.) in the $3\times5$ central \emph{spaxels} of 1.4($\pm$0.4). This
is consistent with C2 being associated with material expelled from the central
SSCs while C1  is associated with a shell around the SSCs.

%
% For built-in environments use
%
%\begin{theorem}
%Theorem text goes here.
%\end{theorem}
%
%\begin{definition}
%Definition text goes here.
%\end{definition}
%
%\begin{proof}
%\smartqed
%Proof text goes here.
%\qed
%\end{proof}
%
\begin{acknowledgement}
Based on observations carried out at  the European Southern
Observatory, Paranal (Chile), programme 078.B-0043(A). This paper uses
the plotting package \texttt{jmaplot}, developed by Jes\'us
Ma\'{\i}z-Apell\'aniz,
\texttt{http://dae45.iaa.csic.es:8080/$\sim$jmaiz/software}.
%This research made use of the NASA/IPAC Extragalactic 
%Database (NED), which is operated by the Jet Propulsion Laboratory, California
%Institute of Technology, under contract with the National Aeronautics and Space
%Administration.
This work has been funded by the Spanish PNAYA, projects AYA2007-67965-C01 and, 
C02 as well as from  CSD2006 - 00070  "1st Science with
GTC"  from the CONSOLIDER 2010 programme of the Spanish MICINN. 
\end{acknowledgement}
%
%\section*{Appendix}
%\addcontentsline{toc}{section}{Appendix}
%
%
%When placed at the end of a chapter or contribution (as opposed to at the end of the book), the numbering of tables, figures, and equations in the appendix section continues on from that in the main text. Hence please \textit{do not} use the \verb|appendix| command when writing an appendix at the end of your chapter or contribution. If there is only one the appendix is designated ``Appendix'', or ``Appendix 1'', or ``Appendix 2'', etc. if there is more than one.

%\begin{equation}
%a \times b = c
%\end{equation}

\bibliographystyle{spphys}
\bibliography{monrealibero_mybib}
\end{document}